\documentclass[preprint,sort&compress,12pt]{elsarticle}

\usepackage{amssymb}
\usepackage{natbib}
\usepackage{amsmath,amsthm,bm,mathrsfs}
\usepackage{color}

 \newtheoremstyle{theorem}{6pt}{6pt}{\rm}{}{\sffamily}{ }{ }{}
 \theoremstyle{theorem}

 \newtheoremstyle{algorithm}{6pt}{6pt}{\rm}{}{\sffamily}{ }{ }{}
 \theoremstyle{algorithm}

 \newtheoremstyle{lemma}{6pt}{6pt}{\rm}{}{\sffamily}{ }{ }{}
 \theoremstyle{lemma}

\newtheoremstyle{case}{6pt}{6pt}{\rm}{}{\sffamily}{. }{ }{}
 \theoremstyle{case}

 \newtheoremstyle{statement}{6pt}{6pt}{\rm}{}{\sffamily}{ }{ }{}
\theoremstyle{statement}

 \newtheoremstyle{corollary}{6pt}{6pt}{\rm}{}{\sffamily}{ }{ }{}
 \theoremstyle{corollary}

  \newtheoremstyle{definition}{6pt}{6pt}{\rm}{}{\sffamily}{ }{ }{}
 \theoremstyle{definition}

\newtheoremstyle{example}{6pt}{6pt}{\rm}{}{\sffamily}{ }{ }{}
\theoremstyle{example}

\newtheoremstyle{remark}{6pt}{6pt}{\rm}{}{\sffamily}{ }{ }{}
\theoremstyle{remark}

\newtheoremstyle{approximation}{6pt}{6pt}{\rm}{}{\sffamily}{ }{ }{}
\theoremstyle{approximation}

\newtheoremstyle{scheme}{6pt}{6pt}{\rm}{}{\sffamily}{ }{ }{}
\theoremstyle{scheme}

\newtheoremstyle{Algorithm}{6pt}{6pt}{\rm}{}{\sffamily}{ }{ }{}
\theoremstyle{Algorithm}

\newtheoremstyle{Assumption}{6pt}{6pt}{\rm}{}{\sffamily}{ }{ }{}
\theoremstyle{Assumption}

\newtheoremstyle{proposition}{6pt}{6pt}{\rm}{}{\sffamily}{ }{ }{}
\theoremstyle{proposition}

\newtheoremstyle{hypo}{6pt}{6pt}{\rm}{}{\sffamily}{ }{ }{}
 \theoremstyle{hypo}

  \newtheoremstyle{Step}{6pt}{6pt}{\rm}{}{}{ }{ }{}
 \theoremstyle{Step}

\journal{Physica A}

\begin{document}

\begin{frontmatter}

\title{%Transparency of a two-level atom near an isotropic photonic band edge
 Susceptibility of a two-level atom near an isotropic photonic band edge: transparency and band edge profile reconstruction}

\author{G. A. Prataviera}
\ead{prataviera@usp.br}
\author{M. C. de Oliveira} 
\ead{marcos@ifi.unicamp.br}
\address{Departamento de Administra\c{c}\~ao, FEARP, Universidade de S\~{a}o Paulo, 14040-905, Ribeir\~{a}o Preto, SP, Brazil}
\address{Instituto de F\'{\i}sica Gleb Wataghin, Universidade Estadual de Campinas, 13083-859, Campinas, S\~{a}o Paulo, Brazil}

\begin{abstract}
{We discuss the necessary conditions for a two-level system in the presence of an isotropic band edge to be transparent to a probe laser field. The two-level atom is transparent whenever it is coupled to a reservoir constituted of two parts - a flat and a non-flat density of modes representing a PBG structure. A proposal on the reconstruction of the band edge profile from the experimentally measured susceptibility is also presented.}
\end{abstract}


\end{frontmatter}



%this files contains Theorem styles based in IMA JOURNALS
%\input standard.tex

%%%%%%%%%%%%%%section A%%%%%%%%%
\section{Introduction}
%%%%%%%%%%%%%%%%%%%%%%%%%%%%%

The investigation of optical properties of atoms coupled to
dissipative environments with a structured density of modes
has been a topic of active research over the years \cite{klepner1,lmg,lzm,klepner2,kurizki,prata1,ref8,poizat}. Of particular interest is the discussion of atoms (impurities) embedded in two or three-dimensional periodic dielectric structures, known as photonic crystals \cite{yablotonovitch,ho,jw1,jw4,moss,nabiev,jq2,kofman,jw5,jw6,tarhan,zhu,paspalakis,zhu2,ref9,ref7,lai,jw3,dalton,huang,hsieh,cheng,yang2,yang1}, since they  allow control
over the electromagnetic density of modes and  the spatial modulation of narrow-linewidth (high-{\it Q}) modes, in both microwave and optical regimes \cite{ref9}. When these structures are used to create one or several forbidden frequency bands they allow control or complete suppression of spontaneous emission, as well as absorption from those embedded impurities \cite{jw1,moss,kofman,jw5,jw6,zhu,paspalakis,ref9,ref7,lai,jw3,dalton,huang,hsieh,cheng,yang2,yang1}. It was particularly relevant the early observation that a two-level atom embedded in a PBG \cite{jw1,jw4,jw5,jw6}  could retain some population in the upper level, even when the transition frequency was in the transmitting band, being the final state a dressed state of the atom with a localized field
mode, which lies in the forbidden band. More recently the attention has been shifted to quantum dots embedded in photonic crystals where each individual quantum dot can be seen as an ``artificial atom'' \cite{vucko,john,nano}. 
The important feature in any of those situations above is that the ``atom''  placed in such a structure
interacts with the field modes in the propagating frequency band and in the
forbidden photonic band gap (PBG) as well, giving rise to many interesting
coherent phenomena such as the possibility of controlling non-markovian decay \cite{kurizki,dalton}, localization of superradiance \cite{jw5}, quantum interference effects in spontaneous emission \cite{zhu,huang}, transparency to a probe field \cite{paspalakis}, and squeezing in the in-phase quadrature spectra \cite{lai}.
%More recently the spontaneous emission of two, three, four and five level atoms embedded in a PBG structure was studied %\cite{zhu,zhu2,zhu3,ding,yang2,yang1}, showing the influence of interference between transitions in the population trapping. 

%The majority of contributions regarding
%radiative properties consider only spontaneous emission, with only a few exceptions treating absorptive and dispersive properties. 
The majority of contributions regarding
radiative properties consider only spontaneous emission of two, three, four and five level atoms embedded in a PBG structure \cite{zhu,zhu2,zhu3,ding,yang2,yang1}, with only a few exceptions treating absorptive and dispersive properties \cite{paspalakis,tajalli}. As an important example of this last case, the absorption and dispersion properties of a $\Lambda$-type atom decaying spontaneously near the edge of a PBG was studied \cite{paspalakis}. It was
pointed out, within an isotropic PBG model, that the atom can become transparent to a probe laser
field, even when other dissipative channels  are present, suggesting that many
surprising effects in the absorption and dispersion of atoms embedded in such structures can appear. Most of those effects were considered inside model systems composed by three or more levels \cite{ref7,kurizki,paspalakis, ref8, ref9, zhu, zhu3,ding,yang2,yang1}, while they were not proved to be strictly necessary.

Pursuing this line we revisit the problem of transparency of an atom placed near an isotropic band edge \cite{paspalakis}, but consider the minimal situation of transitions between two-levels only. We show that for it to be transparent to a weak driven field, the two-level atom must be coupled to a reservoir constituted of two parts - a flat and a non-flat density of modes representing a PBG structure. Transparency is therefore an inner property of the
reservoir engineering. As a side result of this approach we consider the related inverse problem considered in 
\cite{nabiev,PhysRevLett.107.167404,pinkse} on the possibility to obtain information about the band edge profile 
from two-level temporal decay in such structure. Here we show that is also possible to reconstruct 
the band edge characteristics directly from the experimentally measured susceptibility.

This paper is organized as follows. In Section (\ref{s2}), we present the model considered and its stationary solution. In Section (\ref{s3}), the linear susceptibility is evaluated, and two models of isotropic band gap structures are analyzed. In Section (\ref{s4}), is discussed how to reconstruct the band edge characteristics from the experimentally measured susceptibility. Finally, in Section (\ref{s5}) we conclude the paper.
 
\section{Model}{\label{s2}}
The system considered here is a two-level
atom with excited and ground state $\left|1\right\rangle$ and
$\left|0\right\rangle$, respectively and with
transition frequency $\omega_0 $. The atom is probed by a weak electric field
with frequency $\omega $ detuned from $\omega_0 $ by $\delta=\omega - \omega_0$. The decay of the excited state
is due to a coupling with vacuum modes described by a collection
of harmonic oscillators with frequencies $\omega_m $. In the rotating wave
approximation and in the interaction picture the Hamiltonian of the system is
given by
\begin{equation}
\label{1}
H=\left(\Omega e^{i\delta t}\left|0\right\rangle\left\langle 1\right| +H.c.\right)
+\sum_{m} \left(g_m e^{i(\omega_{m}-\omega_0)t}b^{\dagger}_m\left|0\right\rangle\left\langle 1\right| +H.c.\right),
\end{equation}
where $\Omega=-\mu_{10} E_o$ is the Rabi frequency, $\mu_{10}$ is the atomic electric dipole moment, and $E_o$ is electric field 
amplitude. The $g_m$ represents the coupling between
the atom and the vacuum modes and $b^{\dagger}_m$ and $b_{m}$
are the creation and annihilation operators for excitations in the reservoir,
with $m={\lambda,{\bf k}}$ indicating a photon state with polarization $\lambda$ and momentum ${\bf k}$. For sake of simplicity we assume $\Omega$ and $g_m$ as real. In the period of time $t$ the state of total system, {\it atom} + {\it reservoir modes}, can be written
as a superposition
given by
\begin{equation}
\label{2}
\left|\psi (t)\right\rangle = a_0(t)\left|0,\{0\}\right\rangle+a_1 (t)e^{-i\delta t}\left|1,\{0\}\right\rangle + \sum_{m} \alpha_m (t)\left|0,\{m\}\right\rangle.
\end{equation}
The coefficients $a_0(t)$ and $a_1(t)$ are the probability amplitudes to find
the atom in the ground and excited state and the photon reservoir in the vacuum state, respectively, while the coefficient $\alpha_{m} (t)$ gives the probability amplitude to find the atom in the ground state and a single photon in the state $m$ of the vacuum modes. Substituting Eq. (\ref{2}) into the Schr\"odinger equation containing the hamiltonian (\ref{1}) and projecting into each state at the right-hand-side of Eq. (\ref{2}) gives the following equations of motion for the time dependent coefficients $a_{l}$ and $\alpha_{m}$,
\begin{equation}
\label{a0}
i\dot{a}_0(t)=\Omega a_1(t),
\end{equation}
\begin{equation}
\label{a2}
i\dot{a}_1(t)=\Omega a_0 (t) -\delta a_1(t)+\sum_{m}g_{m}e^{-i(\omega_m-\omega_0-\delta)t}\alpha_{m}(t),
\end{equation}
\begin{equation}
\label{a3}
i\dot{\alpha}_{m} (t) = g_{m}e^{i(\omega_m-\omega_0-\delta)t}\, a_1(t).
\end{equation}
Integrating Eq. (\ref{a3}) and eliminating the vacuum amplitude in the equations for $a_0(t)$
and $a_1 (t)$ it follows that
\begin{equation}
\label{3}
i\dot{a}_0(t)=\Omega a_1(t), 
\end{equation}
\begin{equation}
\label{3b} 
i\dot{a}_1(t)=\Omega a_0 (t) -\delta a_1(t)-i\int_{0}^{t}K(t-t')
a_1 (t')dt',
\end{equation}
where the kernel, $ K(t-t')$, is given by
\begin{equation}
\label{5}
K(t-t')=\sum_m g_m^2e^{-i(\omega_m-\omega_0-\delta)(t-t')}.
\end{equation}

All the information about the reservoir is contained in the kernel above,
which will be dependent on the frequency distribution of the vacuum modes.
As we are interested in effects strongly dependent on the reservoir modes
distribution we keep the integro-differential equation (\ref{3b}) without any approximation. 
The procedure to solve Eqs. (\ref{3}) and (\ref{3b}) is straightforward by Laplace transform, which for the initial conditions $a_{0}(0)=1$ and $a_{1}(0)=0$ results in
\begin{equation}
\label{lp1}
V_{0}(s)=\frac{s-i\delta+G(s)}{s\left[s-i\delta+G(s)\right]+\Omega^2},
\end{equation}
\begin{equation}
\label{lp2}
V_{1}(s)=\frac{-i\Omega}{s\left[s-i\delta+G(s)\right]+\Omega^2},
\end{equation}
where $V_0(s)$, $V_1(s)$ and $G(s)$ are the Laplace transforms of $a_0(t)$, $a_1(t)$ and
$K(t-t')$, respectively.
We must only assume two hypothesis to proceed with our calculations. The first
one is about the reservoir - its memory function must allow the
atomic system to reach a steady state, since we are interested in equilibrium
properties at this regime. Secondly, we are interested in the linear atomic response to the external field, so the coupling between field and atom is considered to be weak in such a way that $a_{0}(t)\approx1$
for all times. This means that we are considering a perturbative solution for $a_{1}(t)$, which is linear in the driven field amplitude. This approximation does not affects considerably the results we are discussing, since the only effect of considering a first order term is to neglect power broadening and saturation effects due the laser field intensity \cite{boyd}. In first order in $\Omega$, Eq. (\ref{lp2}) simplifies to 
\begin{equation}
\label{6}
V_{1}(s)= \frac{\Omega}{s[is+\delta+iG(s)]},
\end{equation}
and the steady state solution is obtained by the limit
procedure \cite{barnett}
\begin{equation}
\label{7}
a_{1}(t \rightarrow \infty)=\lim_{s \rightarrow 0}sV_{1}(s)=\frac{\Omega}{\delta+iG(0)},
\end{equation}
where we assumed that the memory function is in such a way that the
limit $s\rightarrow  0$ for $G(s)$ shall exists.

\section{Susceptibility}{\label{s3}}
The induced polarization due the applied external field is given given by 
\begin{equation}
P(t)=\int_{0}^{\infty} \tilde{\chi} (t-t^{\prime}) E (t^{\prime})\, dt^{\prime},
\end{equation}
where $\tilde{\chi }(t)$ is the complex susceptibility, whose imaginary and real part are related to the atom absorption and dispersion of energy from the laser field, respectively \cite{boyd}. For a harmonic field $E(t)=E_{o}e^{-i\omega t}+h.c.$ the polarization becomes
\begin{equation}
P(t)= 2 Re \left(e^{-i\omega t} \chi (\omega) E_o\right),
\label{su1}
\end{equation}
where $\chi (\omega)$ is the fourier transform of $\tilde{\chi} (t)$. On the other hand, the atom polarization is obtained as an average of the atomic dipole moment,
\begin{equation}
P(t)=\left\langle \mu (t)\right\rangle=2 Re\left(\mu_{01}\:a_{0}(t)a_1^{*}(t)\right)\approx 2 Re\left(\mu_{01}a_1^{*}(t)\right).
\label{su2}
\end{equation}
From Eqs. (\ref{su1}) and (\ref{su2}), and considering $N$ atoms by unit of volume, we obtain for the susceptibility at the stationary state, the following expression 
\begin{equation}
\label{8}
\chi(\delta)=- N |{\bf \mu}_{01} |^2 \frac{1}{\delta-iG^{*}(0)},
\end{equation}
which contains the information about the environment through the function $G(0)$. 
Here, we assume that the atomic  decay, i.e., the emission rate into the vacuum modes
can be divided into two parts \cite{lmg,nabiev,kurizki,prata1,ref8} -
the response due to the flat modes of vacuum plus a response due to structured
modes. Following this assumption, we can write the kernel as
\begin{equation}
\label{9}
K(t-t')=\frac{\gamma}{2}\delta(t-t') + \tilde{K}(t-t').
\end{equation}
The first term in Eq. (\ref{9}) corresponds to a Markovian
evolution, due to the coupling to
a flat density of modes  where $\gamma=4\omega_{0}^3|\mu_{01}|^2/3c^3$
is the Wigner-Weisskopf decay rate \cite{ww}.
The second term is the non-Markovian counterpart of the
evolution, corresponding to the coupling with the structured density of 
modes imposed by the  photonic crystal. The linear susceptibility now writes as
\begin{equation}
\label{10}
\chi(\delta)\sim - \frac{1}{\delta-i\frac{\gamma}{2}-i\tilde{G}^{*}(0) },
\end{equation}
where $\tilde{G}(0)$ is the limit $t\rightarrow \infty$ for the Laplace
transform of the non-flat part of the kernel, $\tilde K(t-t')$. 
 
\begin{figure}[tdq]
\includegraphics[scale=1.4]{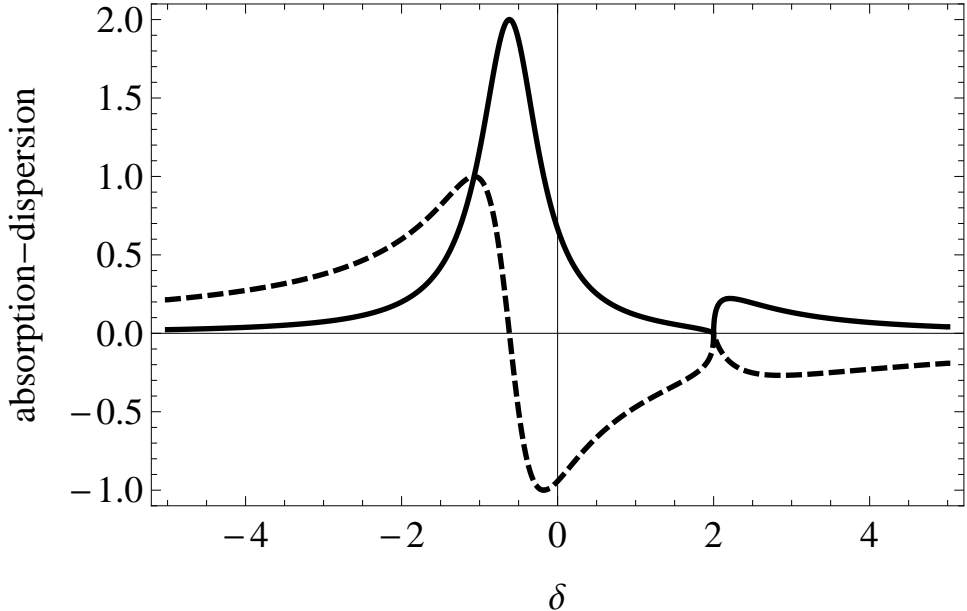}% Here is how to import EPS art
\caption{ Absorption and dispersion as a function of the detuning for the isotropic effective mass model. Solid line: Absorption; Dashed line: Dispersion. We set $\delta_{g}=2$, $\beta =1$, and $\gamma =1$.  The parameters are dimensionless.  }
\label{f1}
\end{figure}

As a first example, we consider that the non-flat part is due to the coupling
to an isotropic photonic band gap where the effective mass dispersion relation is
$\omega_{{\bf k}}=\omega_g+A(|{\bf k}|-|{\bf k_0}|)^2$, with
$A\approx \omega_g/|{\bf k_0}|^2$ \cite{jw1,jw4,ref9}. In this case the kernel is given by
\begin{equation}
\label{11}
\tilde{K}(t-t')=\frac{\beta^{3/2}e^{-i[\pi /4 +(\delta_g-\delta)(t-t')]}}
{\sqrt{\pi (t-t')}},\,\,\,\, t>t',
\end{equation}
with $\beta^{3/2}=2\omega_{o}^{7/2}|\mu_{01}|^2/3c^3$ and
$\delta_g=\omega_g-\omega_o$.  This model corresponds to a density
of modes given by
$\rho (\omega')=\theta(\omega '-\omega_g)/\pi \sqrt{\omega '-\omega_g}$, where $\theta(\omega '-\omega_g)$ is the Heaviside
function. The non-flat kernel then gives us
\begin{equation}
\label{12}
\tilde{G}(s)=\frac{\beta^{3/2}e^{-i\pi /4}}{\sqrt{s+i(\delta_g-\delta)}},
\end{equation}
and the linear susceptibility reads
\begin{equation}
\label{sus1}
\chi(\delta)\sim
-\frac{\sqrt{\delta_g-\delta}}{\left(\delta-i\gamma/2\right)\sqrt{\delta_g-\delta}+\beta^{3/2}}.
\end{equation}
From Eq. (\ref{sus1}) it can be seen that the susceptibility is zero at $\delta=\delta_g$ (or $\omega=\omega_g$), and the atom is transparent to the probe laser field. To illustrate, in Fig. \ref{f1} we plot $Re(\chi)$ and $-Im (\chi)$, which corresponds to dispersion and absorption, respectively, as function of $\delta$, setting $\delta_{g}=2$, $\beta =1$, and $\gamma =1$. Besides transparency (zero absorption) at $\delta=\delta_{g}=2$, it is also observed a strong deviation from the typical two-level absorption-dispersion curves in the Markovian approximation. Thus, the same transparency phenomenon of a three-level system in a $\Lambda $ configuration  \cite{paspalakis} in the presence of a band edge is also possible for a two-level atom. 

\begin{figure}[tdq]
\includegraphics[scale=1.4]{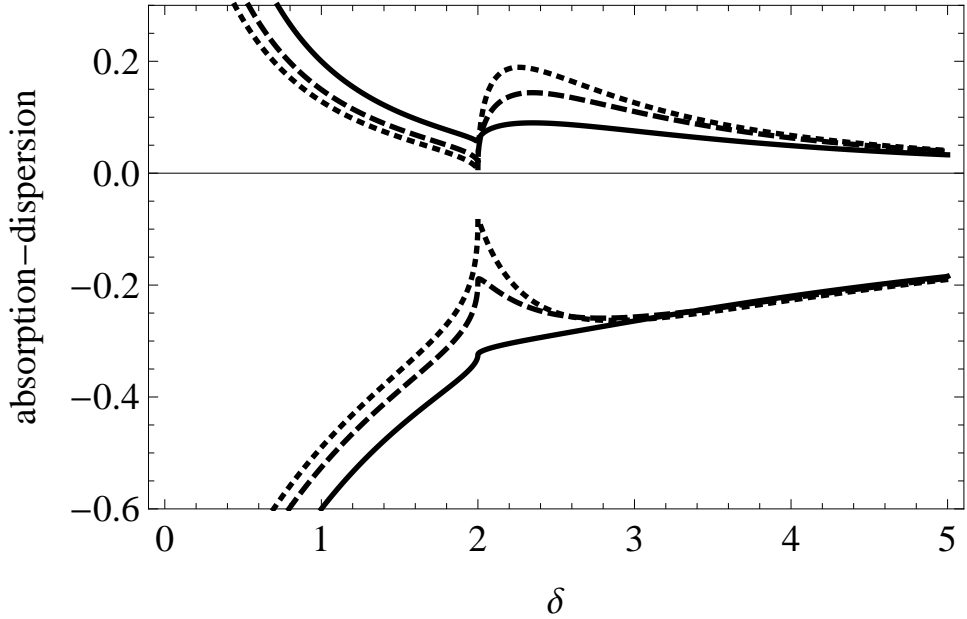}% Here is how to import EPS art
\caption{ Absorption and dispersion as a function of the detuning around $\delta =\delta_g $ for the smoothed one-band isotropic model. Solid line: $\epsilon = 1$; Dashed line: $\epsilon = 0.1$; Dotted line: $\epsilon = 0.01$. We set $\delta_{g}=2$, $\beta =1$, and $\gamma =1$.  The parameters are dimensionless.  }
\label{f3}
\end{figure}

In addition, is worthwhile to consider two extreme cases: (\textit{i}) when the flat part of the density of modes vanish, and (\textit{ii}) when the step-like function describing the band-gap changes in a smooth fashion, as in a real photonic crystal. In the first case, for $\delta < \delta_{g}$, except by a delta function absorption spike at the shifted atomic resonance frequency, the susceptibility given by Eq. (\ref{sus1}) has no imaginary part, and therefore, there is no absorption at these frequencies. For the second case we consider the one band model with 
a smooth density $\rho (\omega')=\sqrt{\omega '-\omega_g}\theta(\omega '-\omega_g)/\pi (\omega '-\omega_g +\epsilon)$, where $\epsilon $ is a smooth parameter to avoid the singularity at $\omega '=\omega_g$ \cite{kurizki,ref8}. The kernel for this density is
\begin{equation}
\label{smo1}
\tilde{G}(s)=\frac{\beta^{3/2}}{ i\sqrt{\epsilon}+e^{i\pi /4}\sqrt{s+i(\delta_g-\delta)}},
\end{equation}
and the susceptibility,
\begin{equation}
\label{sus3}
\chi(\delta)\sim
-\frac{\sqrt{\delta_g-\delta}+\sqrt{\epsilon }}{\left(\delta-i\gamma/2\right)(\sqrt{\delta_g-\delta}+\sqrt{\epsilon })+\beta^{3/2}}.
\end{equation}
In this case the susceptibility is null at $\delta=\delta_g $ in the strict case of $\epsilon =0$. Otherwise, the absorption  and the dispertion curves present a dip at $\delta=\delta_g $ dependent on the value of $\epsilon$. In Fig. \ref{f3} we plot $Re(\chi)$ and $-Im (\chi)$ as function of $\delta$ around $\delta=\delta_g $ considering several values of $\epsilon $. The dip in the line shapes approach zero as the values of $\epsilon $ goes to zero. 

\begin{figure}[tdq]
\includegraphics[scale=1.4]{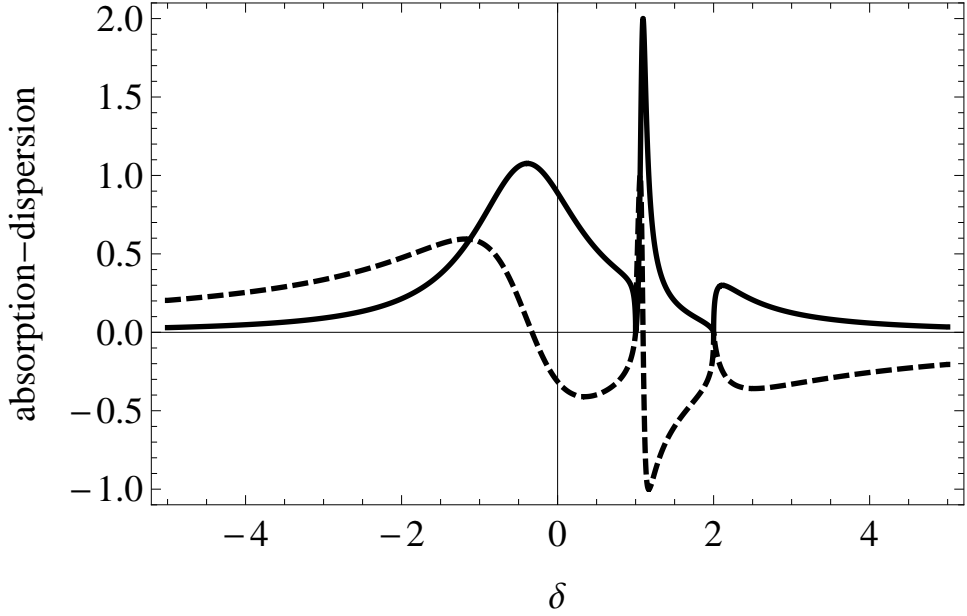}% Here is how to import EPS art
\caption{ Absorption and dispersion as a function of the detuning for the two-band isotropic effective mass model. Solid line: Absorption; Dashed line: Dispersion. We set $\delta_{a}=1$ and $\delta_{b}=2$, $\beta =1$, and $\gamma =1$.  The parameters are dimensionless.  }
\label{f2}
\end{figure}

As a second model for the reservoir we assume that the memory kernel has a non-markovian term due a two-band isotropic effective mass model for the photonic band gap structure. This model corresponds to a density of modes given by \cite{zhu3}
\begin{equation}
\rho (\omega ')=\frac{\theta(\omega_a-\omega ' )}{2\pi \sqrt{\omega_a-\omega ' }}+
\frac{\theta(\omega ' -\omega_b)}{2\pi \sqrt{\omega ' -\omega_b}},
\end{equation}
and resulting  in
\begin{equation}
\label{12b}
\tilde{G}(s)=\frac{\beta^{3/2}e^{i\pi /4}}{2\sqrt{s+i(\delta_a-\delta)}}+\frac{\beta^{3/2}e^{-i\pi /4}}{2\sqrt{s+i(\delta_b-\delta)}},
\end{equation}
where $\delta_a=\omega_a-\omega_o$, and $\delta_b=\omega_b-\omega_o$. Now the susceptibility is given by the following expression
\begin{equation}
\label{sus2}
\chi(\delta)\sim
-\frac{\sqrt{(\delta_a-\delta )(\delta_b-\delta )}}{\left(\delta-i\gamma/2\right)\sqrt{(\delta_a-\delta )(\delta_b-\delta )}
+\frac{\beta^{3/2}}{2}\left(-i\sqrt(\delta_b-\delta )+\sqrt{(\delta_a-\delta)}\right)}.
\end{equation}
In this case the susceptibility is null at $\delta=\delta_a$ or $\delta=\delta_b$,  and the atom becomes transparent at 
two different frequencies. The behavior of dispersion-absorption curves is illustrated in Fig. \ref{f2} where we plot $Re(\chi)$ and $-Im (\chi)$ as function of $\delta$ and setting $\delta_{a}=1$ and $\delta_{b}=2$, $\beta =1$, and $\gamma =1$. Transparency at two different frequencies is observed.

\section{Band-edge profile reconstruction}{\label{s4}}

Since we have developed all the necessary ingredients for understanding the effects of the structured reservoir on the atomic transparency to the probe, we would like to discuss on a rather important related problem, which is the inverse problem of determining the characteristics of the band gap and its profile from
experimental data. As pointed out by Nabiev \cite{nabiev}, the band gap profile by can be determined from the experimental data on the temporal behavior of the atomic spontaneous decay. Indeed, this can also be done by a stationary measure through the susceptibility function. Under the conventional continuum limit for the reservoir mode distribution, and for long times, the Laplace transform for the non-flat contribution can be written as
\begin{eqnarray}
\label{13}
\lim_{s\rightarrow 0}\tilde{G}(s)&=&\lim_{s\rightarrow 0}\int_{0}^{\infty}d\omega^{\prime}\;\Gamma (\omega^{\prime})
\int_{0}^{\infty}d\tau \:e^{-i(\omega^{\prime}-\omega -is)\tau}  \nonumber \\
&=&
\pi \Gamma (\omega)-i\:\mathcal{P}\int_{0}^{\infty}
d\omega^{\prime}
\frac{\Gamma (\omega^{\prime})}{\left(\omega^{\prime}-\omega\right)},
\end{eqnarray}
where $\Gamma(\omega^{\prime})=g^2(\omega^{\prime})\rho(\omega^{\prime})$
represents the product of the coupling with the density of states for the
non-flat sector, and we have used the identity \cite{barnett}
\begin{equation}
\int_{0}^{\infty}d\tau e^{-i(\omega^{\prime}-\omega -is)\tau}=\pi \delta(\omega^{\prime}-\omega -is)-i\mathcal{P}\frac{1}{\omega^{\prime}-\omega -is},
\end{equation}
and $\mathcal{P}$ means principal value. 

For the band edge profile reconstruction it must be noticed that the first term in Eq. (\ref{13}) is exactly taken in
the external field frequency. Thus, when the external field frequency is varied, the reservoir frequency is probed, 
{\it i.e.}, with the variation of the probe field frequency, in fact the band gap frequency distribution is scanned. If one assumes the general expression (\ref{8}) and inverts $\tilde{G}(0)$ as
a function of the measured susceptibility, namely 
\begin{equation}
\label{10b}
\tilde{G}(0) = - \frac{\gamma}{2}+i\left(\delta+\frac{N \left|\mu_{01}\right|^2}{\chi^{*}(\omega)}\right),
\end{equation}
and the band gap profile can be reconstructed.

Remark that the present approach can be applied as well to a broad range of distinct situations, such as in the recent findings on the coupling of atoms trapped in the near field of nanoscale photonic crystal cavities \cite{Tiecke2014}. In this situation the present approach would be useful for probing the cavity density of modes through a susceptibility measurement. 

\section{Conclusion}{\label{s5}

In conclusion, we saw that it is possible to obtain transparency
to a laser probe field on a two-level atom if it is
considered the atomic coupling with a reservoir constituted by flat and
non-flat densities of modes existing in a PBG. We have considered two isotropic band-gap models and analized the linear response to a weak optical field through the susceptibility function. We have also discussed the
possibility of band edge profile reconstruction via the susceptibility function
knowledge. This can be an alternative to the method of reconstruction of the band edge profile involving the measurement of temporal quantities as suggested by
\cite{nabiev} and  implemented with single quantum
dots embedded in a photonic crystal \cite{PhysRevLett.107.167404} (See also Ref. \cite{pinkse} for another method of reconstruction).  In contrast the band
edge profile reconstruction here described can be realized in a steady state
situation. Those findings are particularly relevant for the emerging field of quantum nanophotonics \cite{nano}, as well as in the investigation of nonlinear features with actual atoms trapped in the near field of nanoscale photonic crystal cavities \cite{Tiecke2014}. The measured susceptibility can therefore be a valuable tool for band edge profile reconstruction in those systems.

This work was partially supported by CNPq and FAPESP through the Instituto Nacional de Ci\^encia e Tecnologia em Informa\c c\~ao Qu\^antica (INCT-IQ) and through the Research Center in Optics and Photonics (CePOF).

%%%%%%%%%%%%%%%%%%%%%%%%%%%%%%%%%%%%%%%%%%%%%%%%%%%%%%%%%%%%%

%\acknowledgments{GAP and MCO thank FAPESP (Funda\c {c}\~ao de
%Amparo \`a Pesquisa do Estado de S\~ao Paulo) for financial support.}


\section*{References}
\bibliographystyle{elsarticle-num} 
\bibliography{reference}

\end{document}